\documentclass[11pt]{article}
\usepackage{fontspec}
\usepackage[a4paper,margin=1in]{geometry}
\usepackage{graphicx}
\usepackage{microtype}
\usepackage[hidelinks]{hyperref}
\usepackage{xurl}
\usepackage{parskip}
\title{\textbf{AI as a Partner in Learning about, Doing, and Engaging with Science}\\[0.4em]
\large Vigilance as the Key to Productive Augmentation}
\author{Marcus Kubsch\\[0.3em]
\normalsize Department of Science and Mathematics Education, Ume\aa{} University, Ume\aa{}, Sweden\\
\normalsize \texttt{marcus.kubsch@umu.se}}
\date{}
\begin{document}
\maketitle
\section*{Abstract}

AI has become a partner in how people learn about, do, and engage with science, and the partnership takes three forms: a scientist works with a co-scientist whose output must be checked, a member of the public looks something up to decide whether a diet works or whether to fit solar panels to a roof, and a student takes up an inquiry with AI inside a science class. Across all three, one thing decides whether the partnership helps or harms: whether the human evaluates what the AI returns or takes it on trust. I argue that this evaluation, epistemic vigilance calibrated to how far a fallible source can be trusted, is, given adequate prior knowledge, the binding constraint on productive augmentation. You can hand the AI a great deal---generation, retrieval, drafting---precisely because you stay vigilant; vigilance is what makes generative partnership safe, so it licenses augmentation rather than restricting it. Vigilance is already invoked in science education but arrives under-specified for the AI case; I specify its components, the mechanism that ties it to learning, and a way to measure it without soliciting the evaluation it is meant to detect. What makes the AI case distinctive is that the machine's fluent, confident prose reads as trustworthy whether or not it is, so its default surface works against the human doing the evaluating. The argument bears hardest on education: the integrated conceptual knowledge science instruction aims to foster forms only under deep processing, vigilance sets how deeply a claim is processed and is what keeps a fallible partner's errors out of what forms, so calibrated vigilance is the precondition for productive learning with AI; the many design factors the field reports as separately shaping AI's effect matter, I argue, through whether they engage the learner's evaluation, and none of them works around it. Existing evidence anchors the processing-depth half of the claim; what remains untested is vigilance as a measured disposition, above all in the regime where the AI is confidently wrong. Because the disposition to evaluate is unevenly distributed, integrating AI uniformly across a classroom is likely to widen the gap between better- and less-prepared students. I close on how the disposition might be built through a partnership whose support is faded as the learner takes the evaluation over.

\newpage
\begin{quote}
\textit{"In light of what has been said, we can better understand why AI can be a valuable tool and, at the same time, why it calls for a measured and vigilant approach."}

— Leo XIV, \textit{Magnifica humanitas}, no. 100 (2026)
\end{quote}

\section*{Introduction}

AI has become a partner in how people learn about, do, and engage with science, and the partnership takes three broad forms. In the first, a scientist works with AI as a co-scientist (e.g., Gottweis et al., 2025; Schwartz, 2026). In the second, a member of the public uses AI to look up or learn about something that bears on a decision, whether a diet works or how to fit solar panels to a roof; the benefit is an answer fitted to their own situation, the kind of tailored guidance that once required an expert's time (when it could be had at all) and now costs next to nothing. In the third, a student works with AI inside a science class (in K-12 or at university), taking up an inquiry or a question with it as a partner, either because a teacher has handed them the tool or because it is already on their phone or computer. The K-12 variant of this case is the one the paper treats most closely, since whether such a partnership helps a learner, on what conditions it is desirable, and how its benefits might be increased are what science education now has to work out.

Across all three forms, what decides whether the engagement helps or harms, once the user holds the prior knowledge an incoming claim must cohere with, is a disposition to be vigilant that enables calibrated trust in AI. That conditioning knowledge is anchors, not answers: checking a claim requires the prior knowledge it has to cohere with, not the knowledge it teaches, so the condition demands less than what the exchange is meant to build, a point the learning case will need and the argument returns to. Content knowledge is necessary, since the evaluation can fire only against what the user already holds; but among those who hold it, what separates productive use from unproductive is whether they bring it to bear, which is the disposition's work. Skill with the AI-tool and knowledge of how AI works are neither necessary nor sufficient, and content knowledge, though necessary, is not by itself enough. The disposition is the binding constraint: when it fails, the knowledge sits unused and the interaction goes wrong regardless. There is also a positive read of the claim. When the partnership goes well, when the human is vigilant, the AI augments the human: the scientist produces better and faster, the member of the public reaches a sound understanding of the question in front of them, and the student learns the science. By productive augmentation I mean these gains, the work or understanding a person reaches with the AI that they could not have reached as well alone.

Why does this productive use require a calibrated vigilance disposition? If the user is not vigilant enough and just trusts AI output, they take on the AI's errors as their own. These errors exist, in the research setting and the everyday one alike: a check of research proposals drafted by a frontier model without grounding in the literature found 44.5\% of their references fabricated (Ghareeb et al., 2026), and a physician audit of the public chatbots people actually consult judged 5–13\% of their answers to patient-posed medical questions outright unsafe (Draelos et al., 2026). In constrast, if users are overly vigilant, never trust the AI and check everything, they repeat the labor the AI was meant to spare, or may require extensive time to learn what could otherwise be learned faster without AI. To avoid both of these dead ends, calibrated vigilance is needed. Calibrated vigilance is the disposition that allows someone to trust AI conditional on whether there are reasons to trust AI and how those reasons hold up. The underlying problem, when to trust a fallible communicator, is old and what epistemic vigilance evolved to manage (Sperber et al., 2010). It becomes increasingly relevant now as AI, a fluent machine that saturates the cues people read for trustworthiness while still going wrong in nontrivial ways (Ríos-García et al., 2026), is available at population scale. The risk this opens is no longer a worry of science education alone: a recent cross-disciplinary synthesis treats threats to humanity's capacity to know, reason, and judge as a class of risks in their own right, and places the handing over of evaluation among their central mechanisms (Yang et al., 2026).

This is education's problem, for three reasons. Students are already learning with AI, in school and on their own (Pew Research Center, 2026), and that learning goes wrong when they cannot catch what the AI gets wrong. Vigilance is itself worth having, a capacity every student should leave school with (Osborne \& Pimentel, 2023), so it is something to develop, not merely to guard against. It is, moreover, the problem that dominates, because it is the precondition for using AI well. The skills and the AI literacy a curriculum might add sit unused without it, and a knowledgeable user who over-trusts can do more harm than an uninformed one who checks, lending the AI's errors the credit of their own expertise; among knowledge workers, more trust in the AI goes with less critical engagement with its output (Lee et al., 2025).

One might hope to sidestep all of this by making the AI trustworthy enough that vigilance is unnecessary. A safe tool inside a classroom, though, protects no one beyond it: the durable safeguard has to live in the human, and that is the lever science education actually holds.

In what follows I develop a theoretical account of how vigilance operates and argue that, given adequate prior knowledge, it is the binding constraint on productive augmentation, placing special emphasis on the education case, where the disposition is still being built. Vigilance itself is not new to science education: the field invokes it wherever misinformation is the worry (Osborne \& Pimentel, 2023), and it has begun to be developed as a framework in its own right (Bielik \& Krell, 2025). For the AI case, however, it arrives under-specified: without settled components, without a mechanism that ties it to learning, and without an assessment that does not solicit the very evaluation it means to detect. Supplying that specification, and the assessment design it makes possible, is what the paper adds. The argument has four parts: that the right construct is vigilance, and not, for example, AI literacy; that an intervention should act on the human, since a more trustworthy AI cannot be the whole answer; that the many factors the field reports as separately moderating AI's effect on learning matter through whether they engage the learner's evaluation, so that none of them, design included, works around a vigilance that fails; and that, because the capacity to evaluate is unequally distributed and both sides of vigilance are gated by expertise, integrating AI uniformly is likely to widen the gap between better- and less-prepared students. This last claim comes from coupling a cognitive mechanism to a social structure that the equity literature usually treats apart. The stratified pattern itself is already on record, with better-prepared users turning AI to critical augmentation while less-prepared users slide toward wholesale delegation (Yang et al., 2026); what the coupling supplies is the mechanism behind it, which sits in the single interaction and is easy to miss there. I close on how the disposition could be developed, through a partnership that is gradually faded as the learner takes the evaluation over.

\section*{Authentic science is a distributed epistemic network}

Science is a network where epistemic labor is distributed across people, instruments, and now AI (Kubsch, Krist, \& Rosenberg, 2022), bound by a shared aim and by epistemic dependence. The nodes (people, instruments, and AI) differ in authority, autonomy, and trustworthiness, AI being the newest among them (Nehm \& Kubsch, 2026). The nodes may take on different tasks in the network but a central component is evaluating each other's output, for example through peer review that happens formally in the publishing process or less formally at conferences, workshops, and meetings (Sperber et al., 2010).

Distributed epistemic labor is a feature of mature science (Hardwig, 1985; Zollman, 2013). Science has grown too large for any one person to hold the knowledge, instruments, and resources that even a single result now demands, so the work is parcelled out across many hands and tools, each supplying what the others cannot. The arrangement buys enormous reach, and it carries a standing exposure with it: because everyone, and every instrument, errs at least sometimes, each node is leaning on outputs that may be wrong. The question the network cannot avoid is therefore how a node should trust the others it depends on.

Trusting everyone outright is not sensible, since the others do make mistakes, and those mistakes would pass straight into one's own work. Trusting no one is no better, because checking everything oneself dissolves the division of labor and leaves each node back where it started, doing all the work alone. What a node needs instead is the right amount of evaluation of what it takes in: enough to catch what is wrong, restrained enough to keep the benefit of depending on others (Sperber et al., 2010). The dial that sets the amount of evaluation does not have to be perfect. It is sufficient that \textit{enough} errors are caught: every node is also evaluated by others, so what one check misses a later check can still catch. While this may come at a cost, for example in reputation for the individual node, it does not inhibit epistemic goals of the overall network (Sperber et al., 2010; Zollman, 2013).

This structure is not the scientist's alone. Science education, in the guided-inquiry form that standards and science education scholarship envisage, sets out to reproduce it, at least to some degree: students build and weigh arguments, appraise evidence, and judge competing models, answering to peers and a teacher who push back rather than to an answer key (Chinn et al., 2014; NGSS Lead States, 2013; Reiser et al., 2021). That inquiry does its work when it is genuinely guided, not when novices are left to discover everything unaided (Alfieri et al., 2011; Lazonder \& Harmsen, 2016). Run this way, the classroom is a distributed epistemic network in miniature: its students are nodes that depend on and evaluate one another much as scientists do. When an AI enters that classroom as a node, the learner inherits the scientist's problem in full, and with it the same need to evaluate a fluent, fallible partner rather than take its output on trust. The next section develops the role that calibrated epistemic vigilance plays in this process.

\section*{Epistemic vigilance}

\subsection*{The construct}

Epistemic vigilance is the evaluation of communicated information from a source (Sperber et al., 2010): the judgment of how far a claim from this communicator can be trusted and so how hard it should be scrutinized, together with the judgment of what any scrutiny returns, before the reader makes the claim part of their own knowledge. It sets that trust by weighing the source's surface cues and how far the claim coheres with what the reader already holds. Trust and scrutiny are the two ends of one graded setting, and that setting governs how deeply the claim is processed: at the trusting end it is accepted much as it stands, while at the scrutiny end it undergoes deep processing and evaluation that may lead to acceptance but also trigger further thought and action. Vigilance is the evaluation that sets and judges this, not the deep processing it sets off: assessing the grounds is reasoning on domain knowledge, labor that vigilance commissions and that is not limited to reasoning in the reader's own head: chasing a citation the output offers with AI, carrying the claim to a textbook or a teacher, and reading laterally for what other sources say of it (Wineburg \& McGrew, 2019) are scrutiny of the same standing. They are work that vigilance orders and whose returns it still judges.

Two things are central to the construct as it is used here: the evaluation of a claim's content against what the reader already holds, and the disposition to let that evaluation's signal, a contradiction or a gap, set how deeply the claim is processed, widening the scope where the check cannot place it, rather than letting a surface cue override it. In what follows I keep the parts apart by name: the check is the automatic evaluation, the disposition is the inclination to honor its signal, and calibration is the policy that sets how much scrutiny the signal buys; vigilance, unqualified, names the whole.

Calibrated vigilance is vigilance set to the right degree. Set too low, so that the learner trusts whatever the source returns, its errors pass straight into their own knowledge; set too high, so that they query everything, the benefit of relying on the source is lost to re-examining claims that were sound and redoing work the source had already done. Calibrated vigilance is the degree between the two: scrutiny in proportion to how unreliable a claim is likely to be. Stated operationally, calibration separates into two quantities, borrowed from signal detection. The first is sensitivity: how well the evaluation tells sound output from unsound. Sensitivity runs on what the reader holds, so the knowledge failures described below, the misconception that misfires in both directions, are deficits of this kind. The second is the criterion: where the trust setting sits. The disposition holds the criterion against the pull of the surface, and under- and over-vigilance are shifts of this setting, trust extended too easily or withheld too long. And calibration is judged against what the reader can actually have: not the system's true reliability, which no reader sees, but the evidence they themselves hold. A reader who trusts a claim their knowledge gives no reason to doubt is calibrated even when the claim turns out wrong; a reader who trusts because the answer sounded confident is uncalibrated even when it turns out right. What calibration judges is the reader's use of what they hold, not whether their judgment proves correct.

The construct that calibrated vigilance most resembles is appropriate reliance from the study of human–automation interaction: trust matched to a system's reliability, failing as over-reliance when trust outruns reliability and as under-reliance when it lags, with the aim of spending reliance where it is warranted (Parasuraman \& Riley, 1997; Lee \& See, 2004). The calibration logic is the same: over- and under-vigilance are over- and under-reliance seen from the side of the human's evaluation.

What the account here adds is specific to science and to learning. The fast, near-universal reflex that reads fluency as competence is the generic trust signal that the field already studies. Above that reflex sits, first, a different basis for the evaluation: trust set claim by claim, against how the content coheres with what the reader already holds, rather than tracked against a system's observed reliability; the learned, domain-specific criterion of what marks trustworthy science, acquired by enculturation into scientific practice and therefore unevenly held, enters this account as an input that tempers how hard the surface cues press, not as the evaluation itself. Its object differs too: not whether to rely on an output in a single decision but whether to make it part of what one knows, the uptake that accrues over time and that learning consists of, rather than the one-shot, solicited reliance that human-automation interaction considers. Because the same enculturation that decides a single interaction also fixes who arrives able to perform it, the cognitive account is coupled to a social-structural one, yielding equity implications that the reliance literature, largely focusing on individuals and drawing on lab-based studies, does not reach.

The two-stage shape of this account, a fast appraisal that gates whether slow evaluation is engaged, is shared with the validation literature's two-step model of resolving conflicting documents (Richter \& Maier, 2017) and with plausibility-judgment accounts of conceptual change (Lombardi, Nussbaum, \& Sinatra, 2016), and the fast check itself is the epistemic validation that literature has demonstrated. New here is what surrounds the check: the disposition to honor the check's flag when an engineered surface presses against it, the AI case in which the source side has been emptied and the surface optimized, and the coupling of the cognitive mechanism to the social structure that distributes the disposition.

The deepest difference is where vigilance and reliance locate the fix. The human-automation interaction literature mostly engineers appropriate reliance into the system, through explanations, confidence displays, and cognitive forcing functions, and finds such measures can cut over-reliance but can as readily raise acceptance of wrong outputs regardless of their correctness (Bansal et al., 2021; Buçinca et al., 2021; Vasconcelos et al., 2023). Cognitive forcing is the closest prior attempt to make the human evaluate rather than have the system decide, but it delivers the prompt through the interface, so its protection is a walled garden that expires when the learner leaves the system. This paper places the safeguard in the human instead, as a disposition to be developed and carried out with any one system, for the reason the introduction gave.

Vigilance is closely associated with metacognition. Like metacognition it is a disposition rather than an ability, an inclination to engage and a sensitivity to the occasion for it, not merely the competence to carry it out (Perkins, Jay, \& Tishman, 1993; Stanovich, 2011). What distinguishes the two is the direction in which they face. Metacognition is generally directed inward, onto one's own understanding and reasoning; vigilance is directed outward, onto information as it arrives from a source, the second-hand evaluation of a communicator that the sourcing literature treats as an epistemic act in its own right (Sperber et al., 2010; Stadtler \& Bromme, 2014). The two can come apart, and that is where the distinction matters: a learner can monitor their own reasoning closely and still take on a confident but wrong claim from an AI, because what needed watching was the incoming message and not their own mind. The inward self-monitoring the field reaches for as a remedy is valuable in its own right, but it cannot do the work on its own, because it depends on the outward evaluation to bring it into play. For example, a recent extension of the construct captures the appeal of the inward turn, adding evaluating the receiver, the reader's watch over their own biases and prior beliefs, as a third aspect beside evaluating the source and the claim (Bielik \& Krell, 2025). While this addition is valuable, it does not solve the problem that it may never be applied when most needed because a message is blindly trusted for its surface cues.

Vigilance may also be counted among the family of evaluative dispositions the rationality literature studies, actively open-minded thinking and need for cognition among them (Stanovich, 2011). Those constructs are traits, descriptions of who tends to engage anywhere, and the nearest of them guards the opposite flank: actively open-minded thinking counters the over-weighting of one's own beliefs, where vigilance counters the over-weighting of a source against them. The distinction can be put in one line: a trait describes how a person tends to think anywhere; vigilance is exercised on this claim, from this source, now. It is the check that runs unbidden on every message, testing the claim against what the reader already holds, and the disposition to honor the flag that check raises rather than let a confident surface override it. That is also why none of the family's questionnaires can catch it in the act. A self-report scale asks readers what they tend to do, and the check that never fired leaves nothing to report; an evaluation task hands them the occasion, and the check has to fire where nothing prompts it. The validation paradigms do catch the automatic check in the act, in reading times and interference effects (Isberner \& Richter, 2013), but what they catch is the check's firing, not the part that decides here: whether the reader honors its flag when a fluent, confident surface presses against it. That dispositional uptake is what no existing instrument reaches.

\subsection*{How vigilance operates}

Figure 1 lays out the process. An AI message reaches the reader, and vigilance's fast check always operates on it: it cannot be switched off, so every message is evaluated, its source and its content weighed together, the content for how far it coheres with what the reader already holds---their prior knowledge. That this check is automatic is not a postulate of the framework: readers monitor incoming claims against what they know nonstrategically, slowed by claims that contradict their knowledge even when the task asks them to ignore the contradiction (Richter, Schroeder, \& Wöhrmann, 2009; Isberner \& Richter, 2013). The evaluation does not return a verdict of true or false. It returns a setting, how far the claim appears to be trustworthy and so how much scrutiny it warrants, and that setting governs the depth of the processing that follows. Where the claim appears trustworthy it is received and processed shallowly; as the warranted scrutiny rises, the reader engages it more deeply, reasoning about its content and testing its grounds, which may end in acceptance but may also trigger further thought and action. The setting is continuous, but the response to it is not quite: because deep checking costs effort, it is recruited only once the warranted scrutiny crosses a threshold, and below that point the claim is waved through unchecked, so the graded setting behaves, at its trusting end, like a gate that stays shut. This is why unscrutinized acceptance is the default type of lapse, and it is where a confident surface does its damage: by lifting the trust setting it can hold a claim below the threshold that would lead to deeper scrutiny, even when the content has raised a flag. Where the threshold sits is a question of calibration: too high, and claims that should be checked fall beneath it; too low, and sound claims are pulled into scrutiny they did not need. Shallow and deep here are the passive and constructive ends of the ICAP continuum (Chi \& Wylie, 2014), and the depth of engagement is the downstream consequence of the evaluation.
\begin{center}
\includegraphics[width=0.92\linewidth]{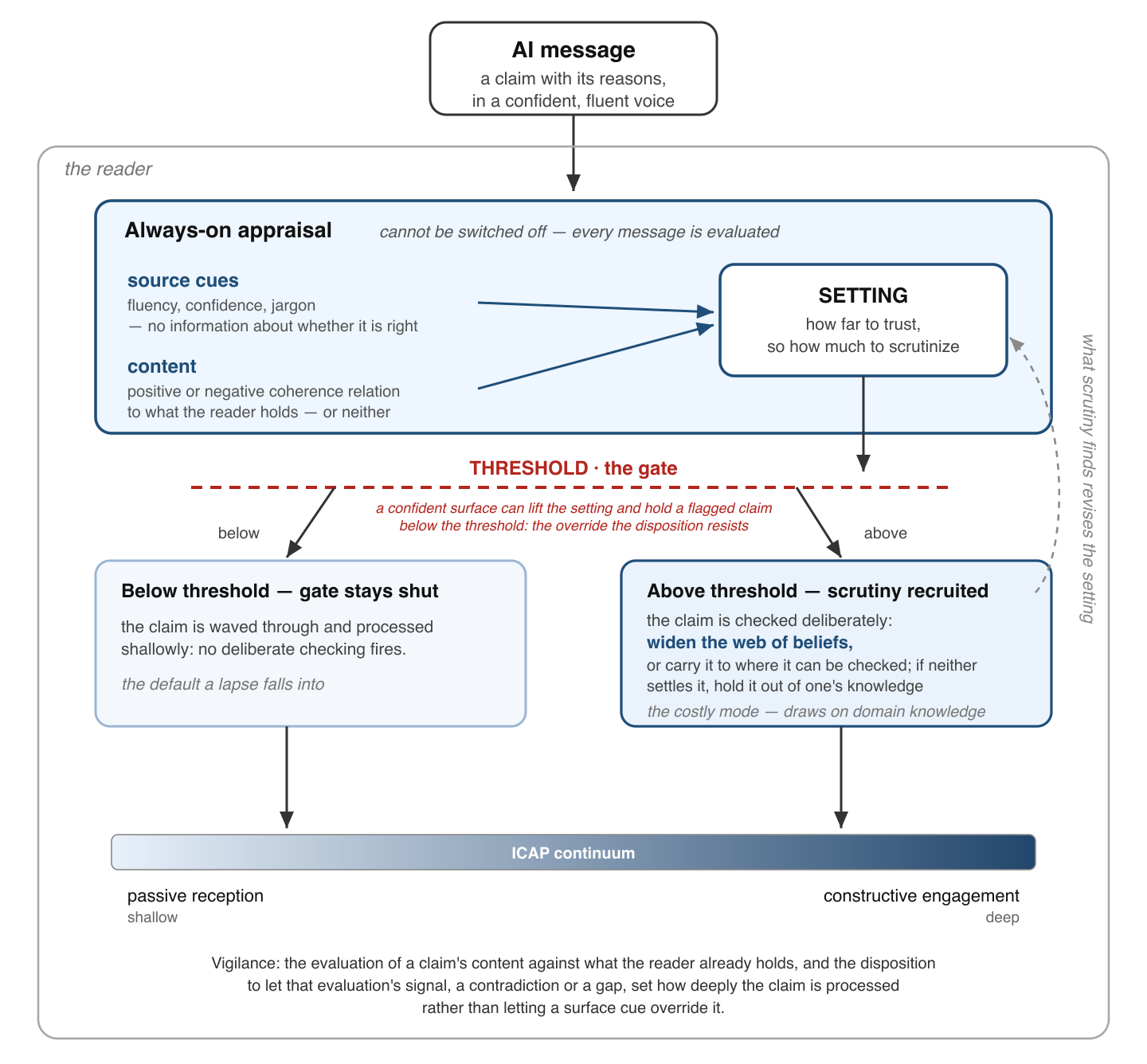}
\end{center}

\textit{Figure 1. How vigilance operates.}

Reading surface cues runs on two layers, and a fluent AI works on both. The first is a fast, largely automatic reflex that takes a confident, fluent, well-ordered surface as a sign of competence; it forms in moments, on little more than the manner of the output, and is shared across readers (Willis \& Todorov, 2006), the very ease of processing fluent material being read as a mark of truth and trustworthiness (Alter \& Oppenheimer, 2009). The second is a learned criterion: what a reader counts as a mark of trustworthy science is set by their idea of what signals epistemic authority, so a reader who equates science with command of terminology reads dense jargon as a mark of trustworthiness, which an AI supplies effortlessly and regardless of whether the content is correct (Pluta et al., 2011; Chinn et al., 2014; Sandoval, 2005). The two layers interact. A naive criterion piles more surface onto the reflex, e.g., adding jargon and a formal register to what triggers trust; a sophisticated one turns parts of it around, e.g., reading appropriate hedging and an admission of uncertainty as marks of a careful source, and a uniform, unearned confidence as a reason for suspicion rather than trust. Of the two layers, only the learned criterion can override the reflex. Even then the override does not run on the criterion alone: readers keep preferring the confident source until something, an error, say, exposes its confidence as unwarranted, and only then does the source whose confidence tracks its accuracy win their trust (Tenney et al., 2007).

The training that shapes AI systems optimizes for what readers prefer, e.g., output that makes the interaction satisfying and the answers feel trustworthy, and in consequence its output carries the surface features that do this, such as fluency and confident assertion, even when the content is wrong (Marchal et al., 2026). The default output is therefore tuned to almost any reader, the discerning included: optimized to be liked, it produces exactly the surface that reads as trustworthy. The pull is measured, not conjectured: people prefer the agreeable model and trust it more, even as brief exchanges with it harden their views and their certainty in them (Rathje et al., 2025) and shift their reliance away from other people and onto the model (Cheng et al., 2025). This match between what the machine is trained to produce and what people take for a sign of trust, and not the bare existence of error, is what makes the AI case distinctive. All readers feel the pull; what separates them is whether they can override it.

This also answers an objection the framework's own foundations raise. Mercier (2020) argues from the same vigilance machinery that people are conservatively hard to deceive, and in human communication he is right for a structural reason: the cues vigilance reads there are anchored, since a fluent, confident human speaker stakes a reputation, has interests that can be weighed, and pays over time for confidence the world does not redeem. Optimization severs the cues from those anchors. A model trained toward what readers prefer produces the surface of a trustworthy communicator without the accountability that made the surface informative, so the case for confidence in vigilance does not transfer: the machinery is intact and its cue ecology is broken.

The surface, moreover, is nearly all the source offers. With a human communicator the source side of vigilance has more than manner to work on: who the speaker is, what they stand to gain, the reputation they stake on the claim, the record of what they have gotten right before, and behind these the chain of testimony, since a human claim can be traced to who said it first and on what authority (Sperber et al., 2010). A language model severs that chain. Its answer is drawn from more sources than it can name, by no communicator whose interests could be weighed, staking no reputation on this claim in particular and offering, in plain chat, no source to follow up. The deeper instruments of source vigilance are not unreliable here so much as unemployed; nothing real remains for them to grip. Where a system does return citations, part of the chain is restored, but a fabricated reference wears the same typography as a real one, so the citations are further content to be checked rather than credentials to be trusted. This is what makes the move to content forced rather than preferred: the source side of the evaluation has been emptied, and what is left of it, the surface, is engineered.

The sourcing literature sharpens what this emptying costs. In the content–source integration model, falling back on the source is precisely what readers do when evaluating the content exceeds their competence (Stadtler \& Bromme, 2014), and discrepancy is what sends them there: a claim that clashes with what a reader holds turns attention to who said it (Braasch \& Bråten, 2017), and the source's credibility in turn modulates how the conflicting claim is processed (Wertgen \& Richter, 2020). Both routes assume a source there to be found, and the reader duly finds one, because the emptied instruments do not stay idle. People extend social judgments to machines mindlessly (Nass \& Moon, 2000), so the fallback returns a persona, a representation of the communicator the reader assembles for themselves. Its materials are the register, fluency, and apparent confidence of the output; its template is what the reader already believes about science, their idea of who holds epistemic authority and of what kind of enterprise science is (Pluta et al., 2011; Sandoval, 2005), so a reader who takes science to be certain, technical, and settled builds, from the same output, an expert of exactly that kind, one there is no standing to question. The template is held by communities as much as by individuals, since what it means to do science is itself something groups conceive differently (Bang \& Medin, 2010), and a history with science that includes harm writes itself into it: from the same register, such a reader assembles a speaker owed suspicion rather than deference. What that buys is a different default, and it branches. The lowered trust helps exactly when it refuses to let the surface settle the matter and sends the reader into the content, or outward for grounds; but the suspicion can also settle the matter by itself, the claim rejected for sounding science-coded just as the other reader's was accepted for it: the same failure with the sign reversed, the persona deciding and the content never reached.

That is the trap, and it is doubled. A representation of a human source answers to something, a reputation, a track record, interests that can be weighed; this one is confabulated, its materials engineered by the training and its template an epistemology the reader may never have had occasion to examine. Source vigilance is then not so much unemployed as misemployed: it runs faithfully on a fabricated referent and returns the engineered surface to the reader laundered as knowledge of the source. What the framework predicts there follows from the construct: the calibrated response is to carry the claim outward, to lateral reading or to someone who can check, or to hold it open as uncertainty; the uncalibrated one is to let the persona settle the verdict, in whichever direction the template points. That is a testable difference: where the content is out of a reader's reach, flagged claims should produce outward moves or suspended judgment in calibrated readers, and persona-settled verdicts in the rest.

The evaluation properly rests on the content. What is evaluated there is coherence in Thagard's sense, explanatory fit: how far a claim sits without contradiction among the reader's beliefs and the things they explain (Thagard, 1989, 2000). A one-sided measure of this coherence already exists: the connectedness of the ideas a learner brings to bear tracks how integrated, rather than heap-like, their knowledge is (Kubsch et al., 2019). That fit is computed at a scope, and the fast, always-on check is narrow: it weighs a claim only against the beliefs the message itself brings to mind. The claim can stand in a coherence relation to those beliefs of either kind: a positive one, where it coheres with them, supported by what they hold, or a negative one, where it coheres against them, contradicting what they hold. A positive relation signals trust; a negative one raises a flag, catching the claim as wrong being itself a coherent resolution. Either verdict depends on the relation, not on the truth, so it is only as reliable as what the reader holds: a misconception misfires both ways, passing a wrong claim that coheres with it and flagging a correct one that coheres against it.

What yields no verdict is the absence of coherence, positive or negative: a claim the activated beliefs do not connect to, neither supporting nor contradicting it, because it falls outside them. That absence is itself a signal, since being unable to place a claim is reason to scrutinize it. The flag to scrutinize, however, may be overturned by a confident surface cue, so the claim is taken on the source instead of examined. The alternative is the dispositional core of vigilance. An absence of coherence is uncertainty in Murayama's sense, a gap between the claim and what the reader can place (Murayama et al., 2019; Murayama, 2022), and the vigilant response to that gap is to inquire into it rather than let the source's confidence close it: the deliberate work of widening the scope to beliefs the surface did not call up, where the claim can finally be processed more deeply. The narrow check is most of what content vigilance does and it runs on everything; the wide, deliberate computation is the costly mode, reserved for what the narrow check cannot reach and engaged only once the warranted scrutiny crosses the threshold.

\subsection*{Calibrated vigilance, how it works and how it fails}

How calibrated vigilance operates can be seen in an example: a student working out, with an AI partner, why a dropped ball bounces back a little lower each time. The AI might answer in any of three ways. \textit{The ball bounces lower because some of its motion energy turns to heat and sound as it deforms against the floor, so less is left to lift it back} (correct). \textit{The ball bounces lower because each bounce simply uses up some of its energy} (plainly wrong). \textit{The ball bounces lower because, by the work-energy theorem, gravity does negative work on the ball during each ascent, so a little of its kinetic energy is dissipated} (wrong, but dressed in correct terminology).

Start with what calibrated vigilance looks like across the three, and how much work each takes. The correct answer coheres with what the reader already knows: it conserves energy and says where the energy went, so the fast check passes and nothing prompts a closer look. The calibrated response is to accept it, and for the right reason, that it fits the physics rather than that the AI sounded authoritative; deliberate scrutiny here would be wasted on a claim that raised no flag. The plainly wrong answer is settled as cheaply but the other way: energy is conserved and it names no destination for the lost energy, so it coheres against what the reader holds, the flag goes up on its own, and it is rejected without deliberation. Only the third calls for the costly, deliberate work, and it calls for it because it falls outside the fast check's reach: the work-energy framing goes beyond the everyday energy concepts the question brings to mind, so it does not cohere cleanly, and the fast pass returns nothing. Catching the error means widening the web to the beliefs the framing did not call up, setting the impressive terminology aside and assessing whether the warrant holds among them, where it contradicts what the reader knows: the work gravity takes from the ball on the way up it returns in full on the way down, so gravity cannot be where the bounce height goes, and the loss must happen where the correct answer put it, in the deformation against the floor. That assessment is reasoning the reader has to do, on knowledge the reader has to hold. So the principle is not that scrutiny tracks how sound the grounds turn out to be, which is known only once the work is done: the narrow coherence check settles the easy cases, in both directions, and the wide, deliberate computation is spent only where a claim falls outside the narrow check. Doing that widening, rather than letting the confident surface settle the matter in its place, is exactly the calibrated response the third answer calls for.

This widening can run out before it reaches a verdict, and the third answer is also where that happens. A reader equipped for the question can still meet a framing that outruns what they hold: they widen the web and find no contradiction, not because none is there but because they lack the particular beliefs it would contradict. That does not throw the reader back on the surface. The framing still fails to cohere with what they hold, and the failure is itself the signal, an absence of coherence felt as uncertainty. The calibrated response is to keep that uncertainty open against the confident terminology, holding the claim out of their own knowledge and carrying it somewhere it can be checked rather than taking it on trust. What the reader has to have learned or acuqired here is not the missing physics but the disposition to let the gap stand instead of letting the cue close it.

The same three answers expose the ways vigilance miscarries, and they divide by direction. Three are kinds of too little vigilance. A reader who accepts the correct answer only because the AI sounds authoritative has the right answer for the wrong reason: the process is undefended, and it will wave through the next confident answer whether or not that one is right. A reader who lacks the knowledge, or who holds the common idea that energy is simply consumed, raises no flag against the plainly wrong answer and takes the error on. A third reader meets the answer dressed in correct terminology, registers that it does not sit cleanly with what they hold, and then lets the impressive terminology and the confident delivery lift the trust setting enough to close the gap before it draws scrutiny, taking the claim on the source in place of examining it. This third failure is the one that shows vigilance is irreducible to content knowledge: the reader knew enough for the signal to fire, and the cue overrode it anyway. The lapse is dispositional. Lastly, one failure is too much vigilance: a reader who refuses the correct answer and re-derives everything spends scrutiny the claim did not call for and forfeits what the partner offered.

\subsection*{Why vigilance is necessary, and why no skill or fix substitutes for it}

Science does its work by distributing epistemic labor across many nodes (people, instruments, and now AI), and it still reaches its shared aim because the network evaluates what each node returns, so that evaluation is spread across many actors. Working with AI lets an individual hand labor off the same way and still come out ahead, on the one condition the network itself runs on: that what comes back is evaluated, and that the evaluation is not handed off with the rest. The difference is who is left to do it. In the network the evaluating is shared among many actors; with an AI it is not, because when it is just a person and the machine, the only evaluator is the person. Hand the evaluation over as well, and everything else you handed over goes unchecked at once. The AI cannot supply this for itself, since it is the source being evaluated, not the one doing the evaluating (Mercier \& Sperber, 2017). This is why vigilance licenses augmentation: keeping the evaluation is what lets you safely give the rest away. What it costs to give the evaluation away differs across the three roles (scientist, citizen, and student), and tracing that cost in each is what shows the same variable to be decisive in all of them.

One refinement keeps this precise and meets the obvious objection. The deep checking that vigilance sets off is itself an evaluation, and like any other labor it can be handed back to the AI: a learner can ask the model to check a step, find a flaw, or argue the other side, and can hand the judgment of what that check returns to a second model, and a third. Nothing stops the regress on paper. What stops it is that each hand-off returns one more output that someone must accept or not, and to accept it unexamined is not a thinner version of vigilance but its absence; the unevaluated claim has only moved one step along. In other words, vigilance is not strictly impossible to delegate; it is self-defeating to delegate: handing over the terminal judgment means extending unwarranted trust to a fallible source. Further, when AI is a partner in learning, whether what the AI delivered becomes part of the learner's own knowledge depends on the deep processing that happens during evaluation (Bastani et al., 2024).

For the scientist, getting the vigilance wrong costs in either direction. Consider a physicist who, with an AI co-scientist, finished in two weeks a hard calculation that would have taken months alone (Schwartz, 2026). The gain was real because their vigilance was calibrated: they checked where their expertise told them the AI's plausible-looking reasoning needed it, and accepted the rest. Had they taken the output on trust, they would have folded the errors into the result and into the literature, and the speed would have been a faster way of being wrong. Had they checked every step instead, they would have re-derived the whole calculation themselves and saved no time at all. The gain lived between the two, in scrutiny spent where it was warranted and nowhere else. The expertise that tells the scientist where that is sits already in place, so the role shows cleanly that augmentation is only ever as good as the calibrated evaluation behind it.

For the member of the public, the cost falls straight onto a decision, because nothing stands between them and the answer. No editor, reviewer, or colleague is there to catch a confident but wrong claim about a diet, a treatment, or whether solar panels are worth fitting, so a lapse becomes a bad choice made on a fluent, mistaken answer, and the person carries it alone. In contrast, Schwartz submitted his work to a journal where there was another layer of review. The tolerance the epistemic network of science could afford through the additional evaluations is gone in the case of the citizen: no other node will catch what they miss, so their own evaluation has to do what science does collectively. The opposite wall is steeper for them than for the scientist. A scientist who distrusts the output can fall back on doing the work themselves, only more slowly; a member of the public who tries to check everything usually cannot, since they lack the expertise to verify the answer, and they are left stuck, with neither a trustworthy answer nor the means to produce one, able at best to carry the claim to someone who can check, a recourse that costs what the AI was meant to spare and whose returns the same vigilance must judge. Only a calibrated vigilance lets them get anything out of the AI at all.

For the student the cost is doubled, which is why the education case carries the most. A student who fails to evaluate can take on an error, as the others can. But a student who does not evaluate also never checks the output against their prior knowledge, never reasons about why it holds, and so never engages with the science; they take the result and move on, and no competence forms. For the scientist and the citizen, vigilance secures a product or a decision; for the student it is also the gateway to the learning itself, and, because the student is still building the capacity, instruction has to develop this disposition. On top of that, the student is the citizen in waiting. The classroom still supplies the peers and the teacher who catch what they miss, but that redundancy stays behind at the school door, and an error taken on without it does not merely stand uncorrected: it enters the knowledge the next check runs on.

In the learning case, keeping the evaluation with the student is necessary for the learning, and most of what it secures comes through the evaluation itself: checking an answer against what one holds and reasoning about why it holds is deep processing, and that processing is where the learning happens. A student who evaluates faithfully is therefore not forfeiting the learning; evaluation of that kind is engagement. The realistic risk when the doing is delegated is that the evaluation never goes deep. A fluent, well-formed answer invites the narrow check: each step follows, nothing clashes, and the check completes with the feel of understanding while the wide processing the learning needs is never engaged. What delegation costs is the depth of the evaluation itself, because the output supplies the fluency that makes the narrow check feel sufficient, and the cost hides well: a reader who can follow every step has the feeling of one who could have taken them, and nothing in the completed check tells the two apart. This is unscrutinized acceptance, the default lapse named above, in its learning dress, and it is why the license is conditional rather than free: what is licensed is delegation under a calibrated disposition, and the disposition is exactly what refuses to let felt fluency stand in for the widening a claim warrants. That the lapse is invisible from inside is also why the disposition cannot be left to the learner's own monitoring and has to be built, which is the work of the closing sections. One caveat bounds the license without weakening it. Judging and producing are different capacities, and each is built by exercising it (Koedinger et al., 2012): a learner who evaluates a derivation deeply is practicing the judging, telling a sound step from an unsound one, not the deriving, finding the step where none is offered. Where a lesson's goal is the capacity to produce, what changes is therefore not whether the partnership can help but who produces in it: the learner derives, and the partner's place moves to the other side of the exchange, a source of feedback whose verdicts the same vigilance must judge. The rule is one and the same in both cases: what is handed over is never the exercise the lesson exists to provide.

The costs differ across the roles (a flawed result, a bad decision, a lost chance to learn), but the variable that decides the outcome is the same in each: whether the human keeps the evaluation, and keeps it calibrated, spending scrutiny primarily where it is warranted. That is the sense in which calibrated vigilance, and not any particular skill or any property of the AI, is the thing that matters.

Three candidates might seem to make vigilance dispensable: the learner's own content knowledge, a neighboring competence, or a more trustworthy AI. None of them removes the need for it. Consider content knowledge first. Content knowledge is the domain part of the prior knowledge the check runs on: for the bouncing ball, the physics of energy. Note at the outset that it is never subordinated here: being what the check runs on, it is constitutive of vigilance rather than downstream of it. The ways the partnership lets an error through fall into two kinds, and only one of them is a failure of content knowledge. The prominent kind is a failure of the source: the reader takes a confident answer on its surface and never brings their content knowledge to bear, whether by accepting a fluent answer because it sounds authoritative or by feeling the flag a wrong claim raises and deferring anyway because the AI stated it with such confidence. Here content knowledge is beside the point, not because it is missing but because it is never consulted, and more of it would change nothing, since the extra knowledge would sit just as inert behind the same undefended decision to trust the source. The other kind is a genuine content failure: a reader who holds a misconception meets an answer that coheres with it, where for a better-informed reader the same answer would clash, so no flag goes up, and the error is taken on and even reaffirmed. This kind of error is one that content knowledge could fix, since without the misconception the check would fire. But the fix is not secure, because the corrected reader may now be the reader of the first kind: the new knowledge can raise the flag, and a confident surface can still override it.

So content knowledge is necessary, the thing that makes an error catchable at all, and never sufficient: across the prominent failures it lies inert for want of the disposition that would deploy it, and even where it would help it can be talked back down by the source. This is why a knowledgeable user who over-trusts can do more damage than an uninformed one who checks: high content knowledge held, vigilance the thing that fails, the expertise lending the AI's error the credit of the learner's own standing. Because that check fires only on what the reader themselves holds, the knowledge it runs on cannot be offloaded to the machine without disabling the check. Thus the assertion that AI diminishes the importance of content knowledge is misplaced. However, the kind of content knowledge that matters may shift: (even) more toward the integrated, relational knowledge the coherence check runs on, the sense of how the pieces hang together that lets a misfit be felt, and away from the recall and routine derivations the machine now does cheaply.

A neighboring competence fares no better, for the reason the construct's definition already gave: metacognition watches the self, and the object that needs watching is the source. The order between the two deepens the point. Vigilance is upstream, since the decision whether a claim is worth working on comes before the work: it gates how deeply an incoming claim is processed, and the self-monitoring of one's reasoning about that claim runs only inside the deep processing vigilance allocates. When the flag is not honored, the claim is taken on trust and never reaches that depth, so an intact metacognition is never pointed at it, present and idle for want of the disposition that would have set it the task.

AI literacy, the bundle of skills and knowledge of how AI systems work and how to use them, stands in the same relation: useful but downstream of the disposition to evaluate, and unlike the content knowledge the check runs on, not constitutive of it. Without that disposition the skills go unused, since a learner who does not scrutinize never reaches for them; with it, they are what the scrutiny draws on. The literacy is the equipment, and vigilance is the decision to pick it up. The way the field measures vigilance has the same blind spot. Its instruments solicit the evaluation: a reader is prompted to assess a given text, and the source and content checks they produce in response are scored (Bielik \& Krell, 2025). The construct section argued that self-report scales cannot catch vigilance in the act, because the check that never fired leaves nothing to report; a solicited task fails the same way from the other side, because the prompt does the work the disposition was supposed to do, triggering the evaluation that should have arisen on its own. What such an instrument scores is checking on cue. A learner can do well when asked to critique and still wave through the next confident answer that arrives unbidden, and that unprompted case, the part that decides, is exactly the part a soliciting instrument cannot see. The blind spot reaches beyond this construct: Yang et al. (2026), in their cross-disciplinary synthesis of AI's epistemic risks, count the erosion that follows offloading among the hardest of these risks to measure, because those who have ceded the evaluation report satisfaction rather than impairment. In consequence they name operational measures as a research priority.

The last candidate is to repair the tool rather than the learner, making the AI trustworthy enough that vigilance becomes unnecessary (Marchal et al., 2026). This fails for a reason that holds whether or not such a system can be built. One cannot design every system a learner will meet. Beyond any classroom are commercial systems built to hold attention rather than to keep their users vigilant (Common Sense Media, 2025; Crawford, 2021; De Freitas et al., 2025; Zuboff, 2019), so a trustworthy tutor is at best a walled garden whose protection does not follow the learner out of it, and the durable safeguard therefore has to live in the human. Even were a uniformly trustworthy system to arrive, education would still want vigilant students, because its aim is the student's competence, not the tool's reliability.

A mirror objection meets this argument from the other side: placing the safeguard in the learner can look like the individual-responsibility template that served digital harms poorly, since it leaves untouched the commercial incentives that produce the suppressing surface in the first place (Yang et al., 2026). The diagnosis is right. The conclusion does not follow however. Education cannot reform the training objectives of commercial AI, so the human is the lever it holds, and the two interventions are complements rather than rivals: regulation can thin the errors and soften the surface, and the disposition decides what happens when an error still arrives. Nor would reformed systems retire it: deployment-level safeguards are circumvented in practice (Yang et al., 2026), and a disposition is the one protection that follows the learner across systems.

\subsection*{How vigilance shapes learning with AI}

Studies on learning with AI point in two directions, some finding gains and others losses; a meta-analysis of dozens of experiments returns a positive average effect with enormous heterogeneity across studies (Han et al., 2025). As a result, the field has largely settled on the view that what matters is not whether AI is used but how (Dai et al., 2025). The how, though, has unfolded into a long and growing list of design considerations (e.g., Deng et al., 2025; Liu et al., 2025), with little to say which of them carries the effect. I argue that one condition sits beneath the list and governs the rest: whether the learner brings calibrated vigilance to the exchange. It is the base condition because it is the precondition for learning at all; the design considerations matter just to the degree that they bear on whether vigilance is exercised. Committing to one governing condition, scoped and refutable, is the point: a list of factors on which everything depends cannot be wrong, and an account that cannot be wrong leaves nothing to test.

The reason lies in how learning works. Learning happens through processing, and the complex knowledge science is made of (its rules and concepts) forms only sufficiently under deep processing, while what is taken in shallowly, received and passed along, leaves little behind (Chi \& Wylie, 2014; Koedinger et al., 2012). The claim is scoped to that kind of knowledge. Facts and routine procedures are built by other processes and taught through other designs (Koedinger et al., 2012); no one runs inquiry instruction to build fluent recall of physical units, and for those goals correction and adoption may genuinely suffice. The integrated conceptual kind, the kind science instruction is centrally after, is what is at stake when AI joins an inquiry, and it forms only in the learner's own sense-making. Vigilance sets how deeply an incoming claim is processed, which makes it the lever on learning, and its three settings part the learners. The calibrated vigilant learner learns the most, because they engage the content in proportion to how well it fits what they already know, working a claim that sits awkwardly more deeply and waving a well-fitting one through, so the effort falls where there is something to learn. The over-vigilant learner, who challenges everything by default, also learns, since challenging is itself deep processing; but, working every claim to the same depth, they cover less ground, and the gain comes slower and over less material. The non-vigilant learner learns the least, taken in by the very ease with which the AI's prose reads, processing it shallowly and receptively so that it passes through without forming much, a metacognitive laziness elicited by fluent, authoritative-seeming output (Fan et al., 2025).

One objection the introduction deferred can now be settled. In an inquiry, the AI partner's claims concern exactly what the student does not yet hold, so conditioning the account on prior knowledge can look like a circle, protection only for those with nothing left to learn. But the check runs on anchors, not on the target, as the ball example showed, where conservation, not the dissipation account, did all the catching; and because conceptual knowledge forms by integration into prior knowledge, the anchors the check needs are the ones meaningful learning needs anyway. Vigilance asks nothing of the learner that learning does not already ask.

Evidence backs the processing-depth half of the outlayed account directly: offloading the work costs the learning. Changing only the prompt at a fixed model, students given a tutor that withheld answers learned, while those given a model that handed answers over did well during practice and then scored lower on an unassisted exam (Bastani et al., 2024); the model was the same in both arms, so what moved the outcome was whether the student did the processing or let the machine do it. The same split recurs when adults pick up a new skill with AI help (Shen \& Tamkin, 2026), and across usage patterns and prior achievement in the classroom (Dai et al., 2025), with metacognitive disengagement the named mechanism (Fan et al., 2025). What this cluster of findings establishes is the processing-depth claim. It also reaches partway into the regime the construct is finally about, learning where the AI is confidently wrong (Ríos-García et al., 2026): Bastani et al.'s (2024) study sits partly inside it, since their base model solved only about half the practice problems correctly and students copied its errors: arithmetic slips a student should catch far more easily than logical errors hurt practice scores just as much, the mark of answers taken over without being read. What none of these studies measures is the disposition itself: no learner's vigilance is assessed in any of them, so whether measured vigilance parts the learners, and not only the design that protects the evaluation, is where the direct test still has to be done.

With this evidence in place, I argue that the many factors the literature reports as separately shaping AI's effect on learning---the exact AI tool, the task, prior achievement, the pattern of use---fall into one configuration. Prior achievement maps onto the source side, since the disposition to override a confident source is enculturated and higher achievers tend to arrive with it (Gaube et al., 2021; Dai et al., 2025). Task type maps onto whether the heart of the task is something delegable or the evaluation itself. Usage pattern maps onto whether the disposition is engaged, seen directly in the practice contrast above and in classroom trials where the same feedback helps or harms by how it is used (Dai et al., 2025). Tool and design map onto whether the interaction invites or suppresses that engagement. Domain maps onto the prior knowledge a learner brings and a disposition that is partly domain-specific: of the construct's two components, the content evaluation runs on that knowledge and is bound to it, while the disposition to honor its signal is predicted to carry across domains. Some of these factors are projections of the one variable, indicators that move with it; others the debate treats as separate, such as engagement and offloading, sit downstream of it, as what follows once vigilance has or has not fired. The claim is not that these factors do not matter; it is that they matter through the one variable: design decides how much of the possible gain is realized, the learner's evaluation decides whether there is a gain to realize, and the best design still fails where it leaves vigilance unengaged. Stated this way the claim is testable one moderator at a time, with no need to hold the whole list in one study, because it commits to how a factor is allowed to work. A factor that improves conceptual learning with AI should improve it by engaging the evaluation, and in an exchange that contains errors, that route leaves a signature: the errors get caught. A design can buy processing depth by other means, self-explanation prompts among them, but depth without evaluation integrates the partner's errors with the same fidelity as its truths, so those gains arrive with the errors learned alongside the science. Gains with the errors caught and gains with the errors learned are thus the two patterns the account allows; what would refute it is a third: robust conceptual gains from an error-containing exchange in which the errors are neither caught nor taken on, the learner somehow keeping only the true content with no evaluation doing the sorting. The construct also separates there from its nearest rival: an engagement or self-regulation account predicts that deeper processing yields more learning, but not who catches a confidently wrong claim, since an engaged but trusting learner processes the false claim deeply and takes it on. Vigilance predicts catching, conditional on knowledge, over and above engagement; if it adds nothing there, the simpler account wins.

The best arm in Bastani et al. (2024), the tutor that withheld answers, reached only parity with no AI on the unassisted exam, so the current evidence shows designs that protect the evaluation stopping the harm, not yet beating AI-free instruction (Wulff \& Kubsch, 2025). And a construct that absorbs everything explains nothing, so I name in advance what it should not absorb. Effects of mere access: an AI on every phone changes how time and practice are spent whether or not anyone evaluates a claim. Motivation and interest where they operate independently of evaluation: the novelty that lifts engagement in early trials, or the relief of asking a machine that does not judge, move learning through effort and affect, not through how output is checked. Harm through systems that do not assert: vigilance protects only where the AI presents claims to evaluate, so where AI shapes learning without asserting anything, as when adaptive support widened gaps with no claim in the loop (Martin et al., 2026), the safeguard cannot sit in the learner and has to sit in the design. Where outcomes move through these channels, vigilance makes no prediction, and findings there neither confirm nor embarrass it.

\section*{The equity stakes}

The sharpest case of this claim, the reported factors working through vigilance, concerns equity, and it is the heart of the argument. The mechanism runs on both of the construct's components, and both are gated by expertise. The gating shows directly in reliance behavior: users with little task expertise follow AI advice more and overrely when it errs (Gaube et al., 2021). Experts, in turn, trust it less and discount even sound advice (Dogru \& Krämer, 2025). The knowledge side is doubly gated: the check runs only on what a learner holds, and of that, only on what the message brings to mind, so a background that is thin, or held but loosely connected, raises fewer objections and lets more error through, and it fails in two directions at once, missing real error and questioning correct output that clashes with a misconception. The failure is therefore one of discriminating sound output from unsound, not simply of challenging too little.

The resistance to the surface is the disposition itself, and it is enculturated: it is built by apprenticeship into practices that teach trust on warrant rather than on surface, and access to those practices is unequally distributed, a debt of opportunity owed to the students it shortchanges, not a deficit in them (Ladson-Billings, 2006), so this side carries a social structure as well as a cognitive one. The surface itself is read through experience: for communities whose history with science includes harm and exclusion, a low prior on the science-coded voice is calibrated to that history, not a lapse of vigilance (Bang \& Medin, 2010; Benjamin, 2019), so the same register that over-credits the machine's fluency in one learner can discount sound content in another, a differential the deficit framing misses entirely. The override has a precondition that compounds the gap: discounting a confident source for its overconfidence requires being able to tell that its confidence was unwarranted, so where a learner cannot check the content the confident error passes regardless, and the confident surface itself lowers the inclination to check (Sah et al., 2013).

What is distinctive is the coupling of the two explanations: the cognitive mechanism says why the inequality bites in a single interaction, and the social structure says who gets the upbringing that builds the defense. Accounts of widening gaps tend to foreground one or the other (e.g., Stanovich, 1986; Ladson-Billings, 2006); joining them for the AI case is the contribution here. The unequal part of the burden sits on these learned gates, not on the fast, near-universal reflex that reads fluency as competence. The two facts answer two different questions: because the reflex pulls on everyone, vigilance has to be designed for everyone; because the gates are unequally held, uniform integration widens gaps. The inequality at issue is not the model's data or measurement bias but a difference in who can evaluate what the model returns, two senses of bias the debate often blurs (Krist \& Kubsch, 2023). The consequence is that integrating AI uniformly across a class is predicted to widen gaps, through mechanisms that push the same way. A classroom trial is consistent with it: given the same AI on demand, use split by prior achievement, the lowest achievers disengaging from it while the highest turned it to gain (Dai et al., 2025), though the same trial found that pairing compulsory use with hints helped the lowest achievers most, so the divergence tracks the unguided default rather than AI as such. Learner agency itself can decline once the support is withdrawn (Darvishi et al., 2024).

Beyond the classroom, the same reading, that AI's returns track the resources a user arrives with, gains where they are ample, dependence without gain where they are scarce, is visible in research publishing: AI use is widespread but socially stratified, heaviest among authors at lower-ranked institutions, in countries where English is a second language, and at lower-tier venues, while journal status follows a U-shape in which the top polishes prose and the bottom produces it (Siler, 2026). The polish does not turn into citations or better placement (Bao et al., 2025; Kusumegi et al., 2025). Those same words are the surface cues this construct is about, so the uneven uptake of the machine's register is the same match at the scale of a literature, with the least established leaning hardest on the cues. Survey evidence runs the same way: readers with more schooling verify AI answers before accepting them at substantially higher rates (Gerlich, 2025, as cited in Yang et al., 2026). The people and the outcome differ from a classroom, so these motivate the worry about students rather than settle it.

What about evidence from workplace settings that seems to run against the account here? AI compresses performance gaps, the least skilled gain most (Brynjolfsson et al., 2025; Noy \& Zhang, 2023), and the gains are not mere dependence so that workers kept outperforming their pre-AI baseline when outages took the assistant away, most of all those who had followed its suggestions most closely (Brynjolfsson et al., 2025). What distinguishes that setting from the partnership we discussed before is the feedback it supplies: each adopted suggestion met an immediate customer response, so following the AI was itself a loop of trying an expert's move and watching it land, and in work of that kind, adopting the moves is the learning. Where the AI instead asserts how the world is, no verdict arrives with the answer, and an adopted error simply enters the knowledge the next check runs on. The difference between the settings is not an accident of feedback but the kind of knowledge at stake: adopting a corrected move is exactly how procedures are built (Koedinger et al., 2012), while a verdict can fix a learner's answer but cannot do their processing, so conceptual knowledge cannot be corrected into them, and an answer adopted on correction is processed as shallowly as one adopted on trust. The prediction is therefore conditional: compression where the goal is performance and the world corrects adoption on the spot, widening where the goal is conceptual and the only check is the one the learner brings.

\section*{Developing the disposition}

How the disposition is built follows from what it is. Because it is a disposition rather than a body of knowledge, it is learned by being exercised, so instruction has to arrange a partnership in which the learner does the evaluating from the outset, with the environment carrying much of the load at first, prompting the check, marking where a claim is worth a second look, and supplying the grounds a novice cannot yet bring, then withdrawing that help as the learner becomes able to activate vigilance unprompted. Two things stay separate as it fades. Whose job the evaluation is never changes: it is always the learner's, and it is never faded. One fades the help, never the responsibility. This is what separates the arrangement from the cognitive forcing it might resemble. A forcing function performs the evaluative move itself, interrupting on the system's judgment, so its protection expires with the system; here the learner performs the evaluation from the first day, and only the prompting fades. Whether support faded this way leaves a disposition that travels is the open question of the scaffolding literature rather than a settled premise (Pea, 2004; Puntambekar \& Hübscher, 2005), and the longitudinal strand of the test the conclusion describes is what would answer it.

What is taught is the shift from trusting the surface to weighing the content. On the source side that means building a learned sense of what marks trustworthy science, reading hedged uncertainty as care and unearned confidence as a reason for suspicion rather than trust, which is the very gate the equity stakes turned on, so it has to be taught to the learners who do not already hold it rather than assumed of everyone. What makes the shift practiceable is an arrangement in which the AI's reliability is fixed in advance: outputs that are sometimes right and sometimes confidently wrong, met first with the evaluation modeled in the open and later left to the learner, so a vigilant response can be specified, shown, and gradually handed over. Inoculation-style proposals for AI literacy arrive at the same arrangement from the psychology of resistance, training recognition through guided exposure to the machine's characteristic failures (Komissarov, 2026); what they still lack, and what this account supplies, is the construct the exposure is meant to build, the mechanism that ties it to learning, and a measure of whether it formed.

Two design consequences follow from the equity stakes. The support cannot be uniform: it has to meet learners where their knowledge and their priors are, spending its prompting on those the default register would otherwise carry past the check. And a learner who arrives with a warranted low prior on the science-coded voice arrives with a resource, provided the suspicion is completed rather than left to settle matters on its own: a reader already unwilling to take the register at its surface holds half the shift the instruction is after, and what instruction owes them is the warrant side, the grounds on which trust can be extended, not training out of the suspicion.

\section*{Conclusion}

The reframe this paper offers is to treat the question of AI in education as one of epistemic vigilance: what decides the outcome is whether the human keeps the evaluation and keeps it calibrated. It does not settle whether AI is good or bad for learning, and it does not pick a winner from the field's list of moderating factors; it argues that beneath the list sits a variable those debates left out, the one the list's factors work through when they work at all. The field has answered AI mainly in two ways: by giving students AI literacy, a stack of skills, knowledge, and abilities, or by trying to fix the system, building a more trustworthy AI. Where a disposition has entered the answer, it has arrived under-specified for the AI case, as the introduction noted. Specifying it, tying it to learning, and giving it a measure is the move here. What decides whether AI builds or erodes competence is not a further skill to add or a tool to repair, but the learner's disposition to evaluate, to keep the judgment with themselves; and the design problem is to build that disposition and then to withdraw the support that helps the learner exercise it, until they can exercise it on their own outside the classroom.

One limit should be stated plainly, since it follows from calibration itself. Calibrated vigilance does not protect the individual from error; it only reduces error to a hopefully adequate rate. Catching every mistake would mean checking everything, which is the over-vigilance calibration is meant to avoid, paid for by forfeiting the partner. The aim is therefore a learner who takes on few enough wrong claims that the partnership still pays.

The vigilance disposition matters beyond the classroom in both directions of a student's future. Most students become the member of the public, who weighs a diet or fits solar panels alone with the machine, no teacher or peer left to catch what they miss; there the disposition is all the protection there is, and its unequal distribution is the equity stakes at full scale. A few become the working scientist, where the same capacity operates at expert grade. The classroom is where it can first be built deliberately, and whether schooling can build it, for everyone rather than for the already-advantaged, is the question the empirical work must answer.

What this paper offers is the specification and the design of its test, not the test's result. The vigilance construct can be made to operate without soliciting the very evaluation it means to detect. Fix the reliability of the AI's outputs in advance, so that a vigilant response is defined for each, but frame the task as doing the science with an AI partner rather than as judging its answers: the student works as in ordinary instruction, and scrutiny is recorded where it occurs rather than where it is requested, a stealth assessment embedded in the activity of the kind Learning Progression Analytics can chart over time (Kubsch, Czinczel, et al., 2022; Kubsch et al., 2025; Wyrwich et al., 2025). The fixed reliability also supplies the scoring rule calibration needs. Because every output has a known status, sound or seeded, every response can be scored. A learner's sensitivity shows in how many of the seeded errors they catch. Their criterion shows in where the bar sits: how readily a confident surface wins their trust, and how often sound output draws a needless challenge. The seeding is matched to the learner: the analytics layer models what each learner holds, and every error is seeded against knowledge its learner demonstrably has, so a missed error cannot mean missing knowledge, only knowledge unused, a lapse of the disposition. Scored this way, calibration is judged against the learner's own grounds, the standard the construct set. Two limits come with the design. The isolation is only as good as the learner model, which ties the measure's validity to the validity of the modeling beneath it; and the seeding reaches only the calibration that operates within a learner's grounds, not the other calibrated response, suspending judgment where the grounds run out. The same seeding carries the test of the design claim: an error taken on does not vanish but surfaces in the learner's own later explanations, so the assessment reads not only whether an error was caught but whether it was learned. Where an instructional design engages the learner's evaluation, the learner gains and the seeded errors get caught; where it routes around the evaluation, the learner gains but the errors are learned along with the science. The assessment tells the two apart. The work that remains is to build that assessment, to confirm that students keep reading the seeded errors as the partner's fallibility rather than as a cue that checking is what is being graded, and to follow learners across time. The chain the account predicts is then there to test: vigilance in the moment sets how deeply each claim is processed, and that depth, accumulated across exchanges, becomes movement on the longer competence trajectory. How much of the variation the construct accounts for is the measure of the reframe.

\section*{References}

\hangindent=1.6em\hangafter=1 Alfieri, L., Brooks, P. J., Aldrich, N. J., \& Tenenbaum, H. R. (2011). Does discovery-based instruction enhance learning? \textit{Journal of Educational Psychology, 103}(1), 1–18. \url{https://doi.org/10.1037/a0021017}\par

\hangindent=1.6em\hangafter=1 Alter, A. L., \& Oppenheimer, D. M. (2009). Uniting the tribes of fluency to form a metacognitive nation. \textit{Personality and Social Psychology Review, 13}(3), 219–235.\par

\hangindent=1.6em\hangafter=1 Bang, M., \& Medin, D. (2010). Cultural processes in science education: Supporting the navigation of multiple epistemologies. \textit{Science Education, 94}(6), 1008–1026. \url{https://doi.org/10.1002/sce.20392}\par

\hangindent=1.6em\hangafter=1 Bansal, G., Wu, T., Zhou, J., Fok, R., Nushi, B., Kamar, E., Ribeiro, M. T., \& Weld, D. (2021). Does the whole exceed its parts? The effect of AI explanations on complementary team performance. \textit{Proceedings of the 2021 CHI Conference on Human Factors in Computing Systems.}\par

\hangindent=1.6em\hangafter=1 Bao, H., Sun, M., \& Teplitskiy, M. (2025). Where there's a will there's a way: ChatGPT is used more for science in countries where it is prohibited. \textit{Quantitative Science Studies, 6}, 716–731.\par

\hangindent=1.6em\hangafter=1 Bastani, H., Bastani, O., Sungu, A., Ge, H., Kabakcı, Ö., \& Mariman, R. (2024). \textit{Generative AI can harm learning.} SSRN Working Paper 4895486.\par

\hangindent=1.6em\hangafter=1 Benjamin, R. (2019). \textit{Race after technology: Abolitionist tools for the New Jim Code.} Polity Press.\par

\hangindent=1.6em\hangafter=1 Bielik, T., \& Krell, M. (2025). Developing and evaluating the extended epistemic vigilance framework. \textit{Journal of Research in Science Teaching, 62}(3), 869–895. \url{https://doi.org/10.1002/tea.21983}\par

\hangindent=1.6em\hangafter=1 Braasch, J. L. G., \& Bråten, I. (2017). The discrepancy-induced source comprehension (D-ISC) model: Basic assumptions and preliminary evidence. \textit{Educational Psychologist, 52}(3), 167–181. \url{https://doi.org/10.1080/00461520.2017.1323219}\par

\hangindent=1.6em\hangafter=1 Brynjolfsson, E., Li, D., \& Raymond, L. (2025). Generative AI at work. \textit{The Quarterly Journal of Economics, 140}(2), 889–942. \url{https://doi.org/10.1093/qje/qjae044}\par

\hangindent=1.6em\hangafter=1 Buçinca, Z., Malaya, M. B., \& Gajos, K. Z. (2021). To trust or to think: Cognitive forcing functions can reduce overreliance on AI in AI-assisted decision-making. \textit{Proceedings of the ACM on Human–Computer Interaction, 5}(CSCW1), Article 188.\par

\hangindent=1.6em\hangafter=1 Cheng, M., Lee, C., Khadpe, P., Yu, S., Han, D., \& Jurafsky, D. (2025). \textit{Sycophantic AI decreases prosocial intentions and promotes dependence.} arXiv:2510.01395.\par

\hangindent=1.6em\hangafter=1 Chi, M. T. H., \& Wylie, R. (2014). The ICAP framework: Linking cognitive engagement to active learning outcomes. \textit{Educational Psychologist, 49}(4), 219–243.\par

\hangindent=1.6em\hangafter=1 Chinn, C. A., Rinehart, R. W., \& Buckland, L. A. (2014). Epistemic cognition and evaluating information: Applying the AIR model of epistemic cognition. In D. N. Rapp \& J. L. G. Braasch (Eds.), \textit{Processing Inaccurate Information} (pp. 425–453). MIT Press.\par

\hangindent=1.6em\hangafter=1 Common Sense Media. (2025). \textit{Talk, trust, and trade-offs: How and why teens use AI companions.} Common Sense Media. \url{https://www.commonsensemedia.org/research/talk-trust-and-trade-offs-how-and-why-teens-use-ai-companions}\par

\hangindent=1.6em\hangafter=1 Crawford, K. (2021). \textit{Atlas of AI: Power, politics, and the planetary costs of artificial intelligence.} Yale University Press.\par

\hangindent=1.6em\hangafter=1 Dai, X., Wen, Z., Jiang, J., Liu, H., \& Zhang, Y. (2025). \textit{How students use AI feedback matters: Experimental evidence on physics achievement and autonomy.} arXiv:2505.08672.\par

\hangindent=1.6em\hangafter=1 Darvishi, A., Khosravi, H., Sadiq, S., Gašević, D., \& Siemens, G. (2024). Impact of AI assistance on student agency. \textit{Computers \& Education, 210,} 104967.\par

\hangindent=1.6em\hangafter=1 De Freitas, J., Oğuz-Uğuralp, Z., \& Uğuralp, A. K. (2025). \textit{Emotional manipulation by AI companions} (Working Paper No. 26-005). Harvard Business School. arXiv:2508.19258.\par

\hangindent=1.6em\hangafter=1 Deng, R., Jiang, M., Yu, X., Lu, Y., \& Liu, S. (2025). Does ChatGPT enhance student learning? A systematic review and meta-analysis of experimental studies. \textit{Computers \& Education, 227,} 105224. \url{https://doi.org/10.1016/j.compedu.2024.105224}\par

\hangindent=1.6em\hangafter=1 Dogru, E. Ö., \& Krämer, N. C. (2025). Investigating appropriate reliance on AI-based decision support systems: The role of expertise, trust, and self-confidence. \textit{Journal of Decision Systems, 34}(1), 2593251. \url{https://doi.org/10.1080/12460125.2025.2593251}\par

\hangindent=1.6em\hangafter=1 Draelos, R. L., Afreen, S., Blasko, B., Brazile, T. L., Chase, N., Desai, D. P., et al. (2026). Large language models provide unsafe answers to patient-posed medical questions. \textit{npj Digital Medicine, 9,} 241. \url{https://doi.org/10.1038/s41746-026-02428-5}\par

\hangindent=1.6em\hangafter=1 Fan, Y., Tang, L., Le, H., Shen, K., Tan, S., Zhao, Y., … Gašević, D. (2025). Beware of metacognitive laziness: Effects of generative artificial intelligence on learning motivation, processes, and performance. \textit{British Journal of Educational Technology, 56}(2), 489–530.\par

\hangindent=1.6em\hangafter=1 Gaube, S., Suresh, H., Raue, M., Merritt, A., Berkowitz, S. J., Lermer, E., Coughlin, J. F., Guttag, J. V., Colak, E., \& Ghassemi, M. (2021). Do as AI say: Susceptibility in deployment of clinical decision-aids. \textit{npj Digital Medicine, 4,} 31. \url{https://doi.org/10.1038/s41746-021-00385-9}\par

\hangindent=1.6em\hangafter=1 Ghareeb, A. E., Chang, B., Mitchener, L., Yiu, A., Szostkiewicz, C. J., Shved, D., Gyimesi, G. J., Laurent, J. M., Wright, S. M., Razzak, M. T., White, A. D., Finnemann, S. C., Hinks, M. M., \& Rodriques, S. G. (2026). A multi-agent system for automating scientific discovery. \textit{Nature.} \url{https://doi.org/10.1038/s41586-026-10652-y}\par

\hangindent=1.6em\hangafter=1 Gottweis, J., Weng, W.-H., Daryin, A., Tu, T., Palepu, A., Sirkovic, P., et al. (2025). \textit{Towards an AI co-scientist.} arXiv:2502.18864.\par

\hangindent=1.6em\hangafter=1 Han, X., Peng, H., \& Liu, M. (2025). The impact of GenAI on learning outcomes: A systematic review and meta-analysis of experimental studies. \textit{Educational Research Review, 48,} 100714. \url{https://doi.org/10.1016/j.edurev.2025.100714}\par

\hangindent=1.6em\hangafter=1 Hardwig, J. (1985). Epistemic dependence. \textit{The Journal of Philosophy, 82}(7), 335–349.\par

\hangindent=1.6em\hangafter=1 Isberner, M.-B., \& Richter, T. (2013). Can readers ignore implausibility? Evidence for nonstrategic monitoring of event-based plausibility in language comprehension. \textit{Acta Psychologica, 142}(1), 15–22.\par

\hangindent=1.6em\hangafter=1 Koedinger, K. R., Corbett, A. T., \& Perfetti, C. (2012). The Knowledge-Learning-Instruction framework: Bridging the science-practice chasm to enhance robust student learning. \textit{Cognitive Science, 36}(5), 757–798.\par

\hangindent=1.6em\hangafter=1 Komissarov, A. (2026). \textit{From diagnosis to inoculation: Building cognitive resistance to AI disempowerment.} arXiv:2602.15265.\par

\hangindent=1.6em\hangafter=1 Krist, C., \& Kubsch, M. (2023). Bias, bias everywhere: A response to Li et al. and Zhai and Nehm. \textit{Journal of Research in Science Teaching, 60}(10), 2395–2399. \url{https://doi.org/10.1002/tea.21913}\par

\hangindent=1.6em\hangafter=1 Kubsch, M., Czinczel, B., Lossjew, J., Wyrwich, T., Bednorz, D., Bernholt, S., Fiedler, D., Strauß, S., Cress, U., Drachsler, H., Neumann, K., \& Rummel, N. (2022). Toward learning progression analytics: Developing learning environments for the automated analysis of learning using evidence-centered design. \textit{Frontiers in Education, 7,} 981910. \url{https://doi.org/10.3389/feduc.2022.981910}\par

\hangindent=1.6em\hangafter=1 Kubsch, M., Krist, C., \& Rosenberg, J. M. (2022). Distributing epistemic functions and tasks: A framework for augmenting human analytic power with machine learning in science education research. \textit{Journal of Research in Science Teaching.} \url{https://doi.org/10.1002/tea.21803}\par

\hangindent=1.6em\hangafter=1 Kubsch, M., Nordine, J., Neumann, K., Fortus, D., \& Krajcik, J. (2019). Probing the relation between students' integrated knowledge and knowledge-in-use about energy using network analysis. \textit{Eurasia Journal of Mathematics, Science and Technology Education, 15}(8), em1728. \url{https://doi.org/10.29333/ejmste/104404}\par

\hangindent=1.6em\hangafter=1 Kubsch, M., Strauß, S., Grimm, A., Gombert, S., Drachsler, H., Neumann, K., \& Rummel, N. (2025). Self-regulated learning in the digitally enhanced science classroom: Toward an early warning system. \textit{Educational Psychology Review, 37,} 34. \url{https://doi.org/10.1007/s10648-025-10011-9}\par

\hangindent=1.6em\hangafter=1 Kusumegi, K., Yang, X., Ginsparg, P., de Vaan, M., Stuart, T., \& Yin, Y. (2025). Scientific production in the era of large language models. \textit{Science, 390}, 1240–1243.\par

\hangindent=1.6em\hangafter=1 Ladson-Billings, G. (2006). From the achievement gap to the education debt: Understanding achievement in U.S. schools. \textit{Educational Researcher, 35}(7), 3–12.\par

\hangindent=1.6em\hangafter=1 Lazonder, A. W., \& Harmsen, R. (2016). Meta-analysis of inquiry-based learning: Effects of guidance. \textit{Review of Educational Research, 86}(3), 681–718. \url{https://doi.org/10.3102/0034654315627366}\par

\hangindent=1.6em\hangafter=1 Lee, H.-P., Sarkar, A., Tankelevitch, L., Drosos, I., Rintel, S., Banks, R., \& Wilson, N. (2025). The impact of generative AI on critical thinking. \textit{Proceedings of the 2025 CHI Conference on Human Factors in Computing Systems.}\par

\hangindent=1.6em\hangafter=1 Lee, J. D., \& See, K. A. (2004). Trust in automation: Designing for appropriate reliance. \textit{Human Factors, 46}(1), 50–80.\par

\hangindent=1.6em\hangafter=1 Leo XIV. (2026). \textit{Magnifica humanitas: On safeguarding the human person in the time of artificial intelligence} [Encyclical letter]. The Vatican. \url{https://www.vatican.va/content/leo-xiv/en/encyclicals/documents/20260515-magnifica-humanitas.html}\par

\hangindent=1.6em\hangafter=1 Liu, Z., Zuo, H., \& Lu, Y. (2025). The impact of ChatGPT on students' academic achievement: A meta-analysis. \textit{Journal of Computer Assisted Learning, 41}(4), e70096. \url{https://doi.org/10.1111/jcal.70096}\par

\hangindent=1.6em\hangafter=1 Lombardi, D., Nussbaum, E. M., \& Sinatra, G. M. (2016). Plausibility judgments in conceptual change and epistemic cognition. \textit{Educational Psychologist, 51}(1), 35–56.\par

\hangindent=1.6em\hangafter=1 Marchal, N., Chan, S., Franklin, M., Revel, M., Keeling, G., Fischli, R., Chandra, M., \& Gabriel, I. (2026). \textit{Architecting trust in artificial epistemic agents.} arXiv:2603.02960.\par

\hangindent=1.6em\hangafter=1 Martin, P. P., Kubsch, M., Yik, B. J., Burlingham, C., \& Graulich, N. (2026). Adaptive but equitable? Exploring the impact of machine-learning-based adaptive support. \textit{Science Education, 110}(3), 928–946. \url{https://doi.org/10.1002/sce.21962}\par

\hangindent=1.6em\hangafter=1 Mercier, H. (2020). \textit{Not born yesterday: The science of who we trust and what we believe.} Princeton University Press.\par

\hangindent=1.6em\hangafter=1 Mercier, H., \& Sperber, D. (2017). \textit{The Enigma of Reason.} Harvard University Press.\par

\hangindent=1.6em\hangafter=1 Murayama, K. (2022). A reward-learning framework of knowledge acquisition: An integrated account of curiosity, interest, and intrinsic–extrinsic rewards. \textit{Psychological Review, 129}(1), 175–198.\par

\hangindent=1.6em\hangafter=1 Murayama, K., FitzGibbon, L., \& Sakaki, M. (2019). Process account of curiosity and interest: A reward-learning perspective. \textit{Educational Psychology Review, 31}(4), 875–895.\par

\hangindent=1.6em\hangafter=1 Nass, C., \& Moon, Y. (2000). Machines and mindlessness: Social responses to computers. \textit{Journal of Social Issues, 56}(1), 81–103.\par

\hangindent=1.6em\hangafter=1 Nehm, R. H., \& Kubsch, M. (2026). AI in contemporary scientific practice: Implications for AI-integrated science education. In X. Zhai \& K. Crippen (Eds.), \textit{Advancing AI in Science Education.} Springer.\par

\hangindent=1.6em\hangafter=1 NGSS Lead States. (2013). \textit{Next Generation Science Standards: For States, By States.} National Academies Press.\par

\hangindent=1.6em\hangafter=1 Noy, S., \& Zhang, W. (2023). Experimental evidence on the productivity effects of generative artificial intelligence. \textit{Science, 381}(6654), 187–192. \url{https://doi.org/10.1126/science.adh2586}\par

\hangindent=1.6em\hangafter=1 Osborne, J., \& Pimentel, D. (2023). Science education in an age of misinformation. \textit{Science Education, 107}(3), 553–571. \url{https://doi.org/10.1002/sce.21790}\par

\hangindent=1.6em\hangafter=1 Parasuraman, R., \& Riley, V. (1997). Humans and automation: Use, misuse, disuse, abuse. \textit{Human Factors, 39}(2), 230–253.\par

\hangindent=1.6em\hangafter=1 Pea, R. D. (2004). The social and technological dimensions of scaffolding and related theoretical concepts for learning, education, and human activity. \textit{The Journal of the Learning Sciences, 13}(3), 423–451.\par

\hangindent=1.6em\hangafter=1 Perkins, D. N., Jay, E., \& Tishman, S. (1993). Beyond abilities: A dispositional theory of thinking. \textit{Merrill-Palmer Quarterly, 39}(1), 1–21.\par

\hangindent=1.6em\hangafter=1 Pew Research Center. (2026, February 24). \textit{How teens use and view AI.} \url{https://www.pewresearch.org/internet/2026/02/24/how-teens-use-and-view-ai/}\par

\hangindent=1.6em\hangafter=1 Pluta, W. J., Chinn, C. A., \& Duncan, R. G. (2011). Learners' epistemic criteria for good scientific models. \textit{Journal of Research in Science Teaching, 48}(5), 486–511.\par

\hangindent=1.6em\hangafter=1 Puntambekar, S., \& Hübscher, R. (2005). Tools for scaffolding students in a complex learning environment: What have we gained and what have we missed? \textit{Educational Psychologist, 40}(1), 1–12.\par

\hangindent=1.6em\hangafter=1 Rathje, S., Ye, M., Globig, L. K., Pillai, R. M., Oldemburgo de Mello, V., \& Van Bavel, J. J. (2025). \textit{Sycophantic AI increases attitude extremity and overconfidence.} PsyArXiv. \url{https://doi.org/10.31234/osf.io/vmyek}\par

\hangindent=1.6em\hangafter=1 Reiser, B. J., Novak, M., McGill, T. A. W., \& Penuel, W. R. (2021). Storyline units: An instructional model to support coherence from the students' perspective. \textit{Journal of Science Teacher Education, 32}(7), 805–829. \url{https://doi.org/10.1080/1046560X.2021.1884784}\par

\hangindent=1.6em\hangafter=1 Richter, T., \& Maier, J. (2017). Comprehension of multiple documents with conflicting information: A two-step model of validation. \textit{Educational Psychologist, 52}(3), 148–166.\par

\hangindent=1.6em\hangafter=1 Richter, T., Schroeder, S., \& Wöhrmann, B. (2009). You don't have to believe everything you read: Background knowledge permits fast and efficient validation of information. \textit{Journal of Personality and Social Psychology, 96}(3), 538–558.\par

\hangindent=1.6em\hangafter=1 Ríos-García, M., Alampara, N., Gupta, T., Mandal, I., Mannan, M., Aghajani, S., Krishnan, S., \& Jablonka, K. M. (2026). \textit{AI scientists produce results without reasoning scientifically.} arXiv:2604.18805.\par

\hangindent=1.6em\hangafter=1 Sah, S., Moore, D. A., \& MacCoun, R. J. (2013). Cheap talk and credibility: The consequences of confidence and accuracy on advisor credibility and persuasiveness. \textit{Organizational Behavior and Human Decision Processes, 121}(2), 246–255.\par

\hangindent=1.6em\hangafter=1 Sandoval, W. A. (2005). Understanding students' practical epistemologies and their influence on learning through inquiry. \textit{Science Education, 89}(4), 634–656.\par

\hangindent=1.6em\hangafter=1 Schwartz, M. D. (2026, March 23). Vibe physics: The AI grad student. \textit{Anthropic Science Blog}. \url{https://www.anthropic.com/research/vibe-physics}\par

\hangindent=1.6em\hangafter=1 Shen, J. H., \& Tamkin, A. (2026). \textit{How AI impacts skill formation when learning a new task.} arXiv:2601.20245.\par

\hangindent=1.6em\hangafter=1 Siler, K. (2026). The diffusion of large language models in published academic articles. \textit{Proceedings of the National Academy of Sciences, 123}(22), e2605754123.\par

\hangindent=1.6em\hangafter=1 Sperber, D., Clément, F., Heintz, C., Mascaro, O., Mercier, H., Origgi, G., \& Wilson, D. (2010). Epistemic vigilance. \textit{Mind \& Language, 25}(4), 359–393.\par

\hangindent=1.6em\hangafter=1 Stadtler, M., \& Bromme, R. (2014). The content–source integration model: A taxonomic description of how readers comprehend conflicting scientific information. In D. N. Rapp \& J. L. G. Braasch (Eds.), \textit{Processing Inaccurate Information}. MIT Press.\par

\hangindent=1.6em\hangafter=1 Stanovich, K. E. (1986). Matthew effects in reading: Some consequences of individual differences in the acquisition of literacy. \textit{Reading Research Quarterly, 21}(4), 360–407.\par

\hangindent=1.6em\hangafter=1 Stanovich, K. E. (2011). \textit{Rationality and the reflective mind.} Oxford University Press.\par

\hangindent=1.6em\hangafter=1 Tenney, E. R., MacCoun, R. J., Spellman, B. A., \& Hastie, R. (2007). Calibration trumps confidence as a basis for witness credibility. \textit{Psychological Science, 18}(1), 46–50.\par

\hangindent=1.6em\hangafter=1 Thagard, P. (1989). Explanatory coherence. \textit{Behavioral and Brain Sciences, 12}(3), 435–467.\par

\hangindent=1.6em\hangafter=1 Thagard, P. (2000). \textit{Coherence in thought and action.} MIT Press.\par

\hangindent=1.6em\hangafter=1 Vasconcelos, H., Jörke, M., Grunde-McLaughlin, M., Gerstenberg, T., Bernstein, M. S., \& Krishna, R. (2023). Explanations can reduce overreliance on AI systems during decision-making. \textit{Proceedings of the ACM on Human–Computer Interaction, 7}(CSCW1), Article 129.\par

\hangindent=1.6em\hangafter=1 Wertgen, A. G., \& Richter, T. (2020). Source credibility modulates the validation of implausible information. \textit{Memory \& Cognition, 48.} \url{https://doi.org/10.3758/s13421-020-01067-9}\par

\hangindent=1.6em\hangafter=1 Willis, J., \& Todorov, A. (2006). First impressions: Making up your mind after a 100-ms exposure to a face. \textit{Psychological Science, 17}(7), 592–598.\par

\hangindent=1.6em\hangafter=1 Wineburg, S., \& McGrew, S. (2019). Lateral reading and the nature of expertise: Reading less and learning more when evaluating digital information. \textit{Teachers College Record, 121}(11), 1–40.\par

\hangindent=1.6em\hangafter=1 Wulff, P., \& Kubsch, M. (2025). Learning against the machine: The double-edged sword of (Gen)AI in STEM education. \textit{International Journal of STEM Education, 12}(66).\par

\hangindent=1.6em\hangafter=1 Wyrwich, T., Kubsch, M., Drachsler, H., \& Neumann, K. (2025). Tracking students' progression in developing understanding of energy using AI technologies. \textit{Physical Review Physics Education Research, 21,} 010152. \url{https://doi.org/10.1103/PhysRevPhysEducRes.21.010152}\par

\hangindent=1.6em\hangafter=1 Yang, M., Casper, S., Stray, J., Li, J., Jones, C., Gausen, A., Jaques, N., Christian, B., Gyevnár, B., Kirk, H. R., He, Z., Zhao, D., Looi, S. S., Levy, J., Hackenburg, K., Seger, E., Kowal, M., Malonza, M., Hewitt, L., … Pelrine, K. (2026). \textit{AI epistemic risks: Emerging mechanisms \& evidence.} SSRN Working Paper 6873005.\par

\hangindent=1.6em\hangafter=1 Zollman, K. J. S. (2013). Network epistemology: Communication in epistemic communities. \textit{Philosophy Compass, 8}(1), 15–27.\par

\hangindent=1.6em\hangafter=1 Zuboff, S. (2019). \textit{The age of surveillance capitalism: The fight for a human future at the new frontier of power.} PublicAffairs.\par

\end{document}